# TanGi: Tangible Proxies for Embodied Object Exploration and Manipulation in Virtual Reality

**Martin Feick**[1,4]   **Scott Bateman**[2]   **Anthony Tang**[3]   **André Miede**[4]   **Nicolai Marquardt**[1]

[1]University College London, London, UK    [2]University of New Brunswick, New Brunswick, Canada
[3]University of Toronto, Toronto, Canada    [4]htw saar, Saarbruecken, Germany
martin.feick@dfki.de, scottb@unb.ca, tonytang@utoronto.ca, andre.miede@htwsaar.de,
n.marquardt@ucl.ac.uk

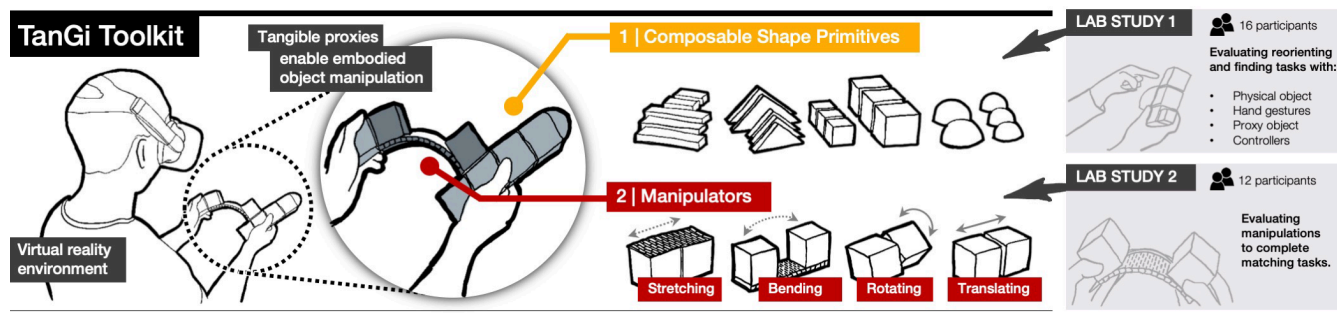

**Figure 1: Visual abstract providing an overview of TanGi elements and the context of our two user studies.**

## ABSTRACT
Exploring and manipulating complex virtual objects is challenging due to limitations of conventional controllers and free-hand interaction techniques. We present the TanGi toolkit which enables novices to rapidly build physical proxy objects using *Composable Shape Primitives*. TanGi also provides *Manipulators* allowing users to build objects including movable parts, making them suitable for rich object exploration and manipulation in VR. With a set of different use cases and applications we show the capabilities of the TanGi toolkit, and evaluate its use. In a study with 16 participants, we demonstrate that novices can quickly build physical proxy objects using the *Composable Shape Primitives*, and explore how different levels of object embodiment affect virtual object exploration. In a second study with 12 participants we evaluate TanGi's *Manipulators*, and investigate the effectiveness of embodied interaction. Findings from this study show that TanGi's proxies outperform traditional controllers, and were generally favored by participants.

## INTRODUCTION
Virtual Reality interfaces will fundamentally change how we design and work with physical objects. VR-based 3D content creation systems allow rapid prototyping of 3D models by using head worn displays and employing 6-DOF controllers. These controllers give designers a type of embodiment in the virtual space, allowing them to move, place and rotate the models using 3D controls. Furthermore, they enable new, intuitive ways to create and engage with 3D objects compared to completing these tasks with traditional 2D interfaces [2, 3, 29, 31, 42, 45].

Despite the improvements offered by 6-DOF controllers to facilitate rapid creation, exploration and manipulation of 3D objects, working with virtual 3D models can still be challenging, because current interfaces are disembodied. For example, a designer creating a new toy relies on controller-based manipulations to move parts of the virtual toy around, and this sort of control-display remapping is cumbersome. The designer cannot feel and easily test out the object through the controllers, and studying how different parts of the toy will behave and react when they are physically manipulated relies on imagination, since controls are not a direct analog for how the toy would really feel.

In this work, we deepen research into how we can give an *embodiment* to virtual objects, by giving them tangible form and moveable parts that match their virtual counterparts. Recent work has highlighted that providing a physical proxy for virtual objects can facilitate interactions [20, 21, 40, 46, 55]. Our work extends these findings, enabling embodiments to be created for virtual objects by providing a toolkit that allows the creation of *tangible proxies* – rapidly built physical stand-ins that approximate key elements of both form and function of a virtual object. Our toolkit, called *TanGi,* enables users to create representations that allow proxy object *manipulations,* such as bending, stretching, and rotating.

The TanGi toolkit provides both *composable shape primitives* (to approximate the size and shape of the virtual objects), and a representative set of *manipulators* (which allow multi-part objects to move in relation to one another through rotating, stretching and bending operators). Figure 1 illustrates a proxy object which is assembled using TanGi primitives, and allows for manipulations.

To evaluate how object embodiments created with TanGi can affect interactions, we conducted two lab studies that explored object exploration and manipulation. The first study showed that for reorientation and finding tasks, embodied proxies offered quicker completion times and physical operations that aligned more closely to people's expectations. The second study showed that participants could use the proxies to more quickly and accurately complete matching tasks required manipulating different parts of a proxy.

This work makes three major contributions: first, we present the conceptual design of TanGi, a toolkit that enables embodied object manipulation in VR; second, through presenting different use cases and applications we show the capabilities and expressive power of the TanGi toolkit. Finally, we show that physically embodied virtual objects enable improved exploration and manipulation on virtual objects.

## RELATED WORK

The HCI community uses the term "embodiment" in a number of ways. In this paper, we refer to embodiment in two ways: first, the proxy object gives physical embodiment to the virtual object; second, how people interact with the virtual object thus becomes embodied since interactions with the object are more direct—manipulations on the physical object are mirrored in the virtual world. Therefore, we situate our work within the context of tangible and embodied interaction research, where research has long focused on the cognitive benefits of using tangibles to interact with computation.

***Tangible Interfaces and Embodied Interaction***. Embodied interaction argues that when people can interact cognitively and physically with information (e.g., through tangible interfaces [26]), people can more fluidly understand the information being manipulated [12]. We have seen, for instance, that tangible interfaces promote natural interaction [42], are faster and more intuitive to use [8], because they benefit from human spatial memory [11].

Recent research has explored how to use tangible real-world objects as physical proxies for virtual models [4, 17, 20, 54]. For instance, Hettiarachchi et al. [20] show how an AR system can automatically identify nearby real-world objects that offer the best physical approximation of a virtual object, to be used as a proxy object. The downside of this approach is that multiple objects with various features need to be nearby, and real-world objects may only roughly match the shape of the virtual counterpart. Other work has shown that such mismatches between physical proxies and virtual models hinder interactions, pointing out that mismatches are most significant for tactile feedback, temperature and weight differences [47]. Importantly, proxy fidelity affects immersion in the virtual environment, performance, and the intuitiveness of interacting with virtual objects; the higher the proxy fidelity, the better the interaction [41].

Providing haptic sensation for virtual models frequently requires unwieldy or bulky hardware. Various devices create different haptic sensations including rendering the shape of physical objects [7, 36, 37], providing force-feedback [19], dynamic weight-shifting [55], or may be used for character animation [28] and object construction tasks [34]. Some haptic devices overcome this with wearables that simulate weight and grasping [10, 43] using electrical muscle stimulation [34, 35]. Robots can provide physical props for a virtual environment [48, 50], and drones can provide haptic feedback for interacting with virtual models [1, 24]. Similarly, shape-changing interfaces are promising, but can be bulky [14] or challenging to fabricate [54]. Haptic feedback has also been shown to improve immersion in virtual reality-based navigation tasks, since the navigation becomes an embodied task [27]. TanGi builds on the idea of embodiments by providing real-world proxies for manipulating virtual models, and extends this idea beyond composable primitives (e.g., Muender et al. [41] use Lego blocks) by adding manipulators that allow the proxies to be multi-part objects that move in relation to one another.

***Toolkit Research & Rapid Fabrication.*** Using real-world proxies to interact with virtual models provides clear advantages [21, 40, 41, 55], but it is largely impractical to have proxy objects for every virtual model. Cheng et al. [9] propose using sparse haptic proxies through a haptic illusion, but this may not be possible for complex shapes. Others have explored how 3D printing enables new opportunities to quickly prototype/build proxies on demand. Mueller et al. [40] use a combination of 3D printing and Lego bricks to build functional proxy objects twice as fast as traditional 3D printing.

Toolkits should aim to minimize the difference between what is possible with the virtual and what is possible with the proxy [30, 46]. Real-world objects have a vast complexity in terms of movable parts; e.g. some are rotatable, bendable, stretchable and translatable. Following these ideas, researchers have also identified manipulations of proxies as an important next step in improving interactions through proxies. The HapTwist [57] toolkit uses unified parts connected via twistable joints. It offers better robustness; however, it does not allow to replicate manipulable object parts. VirtualBricks [5] provides Lego-based proxies that allow for translation and rotation of a proxy.

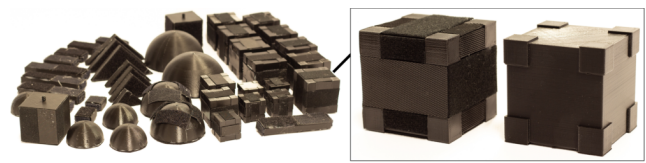

**Figure 2: TanGi Toolkit (left) and 50mm base cubes showing ablated areas and Velcro tape pattern (right).**

Our toolkit share similarities with HapTwist [57] and VirtualBricks [5], but extends the idea of manipulable parts by introducing two new types of manipulations (i.e., variable linear stretching and unidirectional bending). Further, we provide a first evaluation providing clear results that proxies better support exploration and manipulation interactions, when compared to conventional controllers.

**TANGI TOOLKIT CONCEPT AND DESIGN**

Similar to earlier work, we are motivated by the need of providing designers with the ability to rapidly prototype physical proxies that can enable embodied exploration and manipulation. Our approach relies on *composable shape primitives*, which allow rough tangible proxies to be constructed quickly, and *manipulators*, which allow multi-part objects to be composed with moving parts. Together, these enable embodied exploration by matching the tangible proxy to the virtual object, and embodied manipulation by allowing the tangible proxy to control the virtual object. In our vision TanGi is extendable. Thus, designers can create customized composable shape primitives which meet their own requirements, and subsequently they can re-use them.

The TanGi toolkit philosophy was driven by three goals. First, the toolkit should enable rapid iterative prototyping with very quick turnaround (<5 mins). Second, the proxies made with the toolkit should enable exploration of corresponding virtual objects. Third, the proxies should allow people to manipulate the virtual objects.

**Composable Shape Primitives for Embodied Exploration**

Whereas others try to solve the exploration problem by either repurposing real-world objects [20], 3D printing techniques [25, 39, 40] or through robot assemblies [18, 50, 54] our approach relies on *composable shape primitives*, allowing people to create proxies that approximate virtual objects.

In the first version of this toolkit we provide four primitive shapes at three different sizes: cubes, triangles, half-spheres and sticks (Figure 2 left). We decided on these primitive shapes after a formative prototyping phase with foam board. These shapes can be composed into larger composite objects using heavy-duty Velcro tape. As illustrated in Figure 2 (left), the primitives allow us to replicate a variety of basketball-sized objects.

Our implementation relies on 3D printing to fabricate the shapes, and a Velcro-pattern (Figure 2 right) on the cubes that provide a stable base atop which additional shapes can be applied. These proxies can thus be composed of reusable primitives that can be built up and taken apart to represent various virtual objects as necessary. This approach is similar to the often-used *block structures* [5, 40, 44, 51]. Going beyond using traditional brick structures, TanGi can provide a richer set of shapes primitives and can be easily extended with by adding new 3D-printed primitives when necessary.

When combined with a 3D tracker (in our current version, a Vive Tacker) objects composed with TanGi can function as a tangible proxy that can be used to control the movement and orientation of a corresponding virtual object. This allows people to engage in embodied exploration, moving, feeling, reorienting and grabbing approximation of different parts of the virtual object.

*Fabrication.* We designed a basic set of primitive shapes using the CAD software Rhino3D [59] (Version 6 SR14). The models were exported as stereolithography (.*stl*) files, and printed on a Fused Deposition Modeling (FDM) 3D printer using PLA. As visible in Figure 2 (right), the design offers ablated areas of 1mm to accommodate the heavy-duty Velcro tape. All four shapes were fabricated in three different sizes e.g. the cube in 50, 40 and 30mm. Overall, we fabricated 56 objects.

**Manipulators for Embodied Manipulation**

Physical objects have vast complexity such as rotating parts, can be stretched, folded, deformed, bended etc. TanGi provides a representative set of manipulators that allow multi-part objects to move in relation to one another, in an effort to minimized the difference between physical proxies and their virtual counterparts (as suggested by [47]). While the entire range of manipulations that are possible with a physical/virtual object is beyond the scope of this work, we developed TanGi with the goal of incorporating a larger set of representatives of manipulations than has been done in previous work. TanGi manipulators replace the previously described Velcro connectors between shape primitives with new manipulable blocks. Manipulators allow for a movement relationship (i.e. rotation, translation, stretching, bending) between shape primitive to be tracked. These movements can then mapped to the virtual object, allowing parts of the virtual object to be controlled.

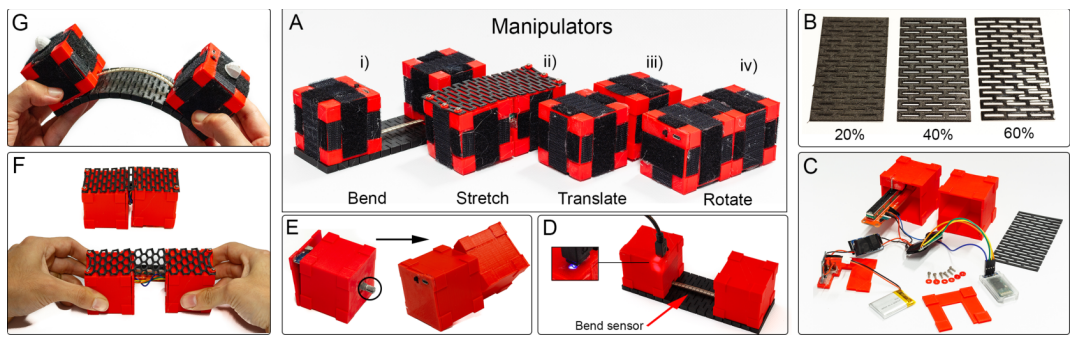

**Figure 3: Clockwise: (A) The four different Manipulators: (i) bend, (ii) variable linear stretch, (iii) linear translation and, (iv) single-axis rotation, all augmented with Velcro tape. (B) Variable stretching patterns. (C) All the components inside a stretching Manipulator including the modular stretching pattern. (D) Charging the battery; when charging, the LED is red - blue when done. (E) shows a rotation manipulator connected using a rotary potentiometer. (F) and (G) show stretching and bending.**

In this first version of the toolkit, we focused on four movement primitives, which we describe below. We expand on variable linear stretching and unidirectional bending, since these are new contributions of our work.

- **single axis rotation**: Enables objects to have rotational parts (e.g. bottle lid) through using a rotary potentiometer.
- **linear translation**: Parts of an object can be moved back and forth in one direction (e.g. linear sliders). This manipulator utilizes a linear potentiometer.
- **variable linear stretching**: Extends linear translation by providing a better sense of how much parts of the object can be translated in order to communicate min/max states. As a result of the increasing amount of force needed to stretch the object (e.g. to cock a crossbow). It uses the same hardware base as the linear translation manipulator; additionally, it utilizes a variable 3D printed stretchable material on top, which provides force-feedback. Following TanGi's modular approach the stretching patterns can be replaced. Thus, users can choose between less stretchable (more force required) or more stretchable (less force required) pattern to create different haptic sensations.
- **unidirectional bending**: Enables objects that have bendable parts, such as a fishing rod. It also naturally communicates min/max states. To achieve this, we use a bend/flex sensor between two distant cubes. Similar to the stretching pattern we utilize a bending pattern between the cubes. Depending on the 3D printed pattern users can create a less/ more bendable object. In the default position the bend manipulator is straight.

*Fabrication and Implementation*. We modified our cube primitive to accommodate all components and parts inside. To fabricate the patterns that allow stretching and bending, we used *thermoplastic polyurethane* (TPU), an elastic 3D printing material. We designed the different patterns in Rhino3D inspired by work on stretchable circuits [16] (Figure 3b). Each Manipulator uses low-cost off-the-shelf hardware components: an hc-06 Bluetooth module, an Arduino Nano 3.x, a voltage converter, a charging unit (chip TP4056), a 3.7V 400mAh Lithium Polymer battery, a switch, wires, resistors and different sensors. Manipulators are self-contained and do not require external power or connection cables to transmit data. The Arduino inside the manipulator continuously executes code for resistive sensing, and sends updates via Bluetooth to the serial port of the VR machine, providing a sampling rate of at least 30 Hz. VR scenes are created in Unity3D. STL models, circuit schematics and the processing code is open-source and can be downloaded from (https://github.com/MartinFk/TanGi).

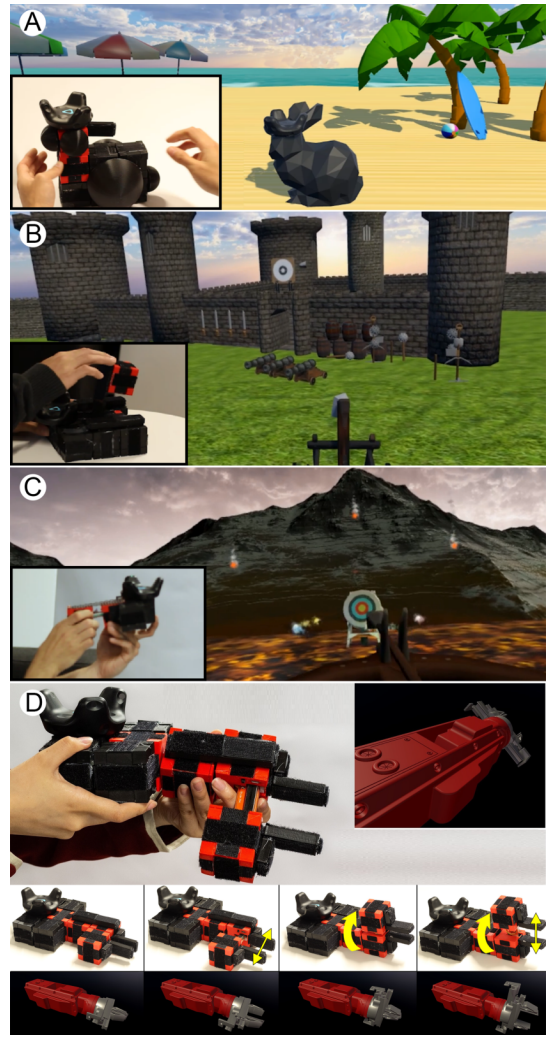

**Figure 4: Example use case applications.**

## EXAMPLE APPLICATIONS & USE CASES OF TANGI

The resulting TanGi toolkit, composed of shape primitives and manipulators, enables users to quickly build a variety of objects allowing complex manipulations. In this section, we illustrate how the toolkit can be used to build several different tangible proxies that represent and control virtual objects. Importing our Manipulator module (a library including all components for serial communication) into Unity3d provides the VR interface for the manipulators.

As our first example, Figure 4a shows a modified *Stanford bunny toy* which can turn its head, using a single-axis-rotation manipulator. Its virtual representation gets rendered accordantly to the physical proxy object. To do so, a user would simply import 3D models for the bunny's head and body, and by attaching the *RotationManipulator* script in Unity, the virtual bunny can now receive rotation updates.

The second example shows a *gameplay catapult* that utilizes a bending manipulator (Figure 4b). Users can move the catapult to the desired location and load it by bending the manipulator. To launch a virtual stone the user releases the cube on the end of the bend manipulator which then accelerates. The max. bend state achieved determines how much force is applied to the stone affecting its trajectory.

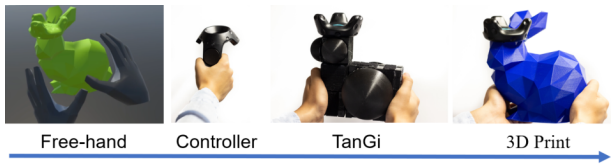

**Figure 5: Different levels of proxy embodiment in Study 1.**

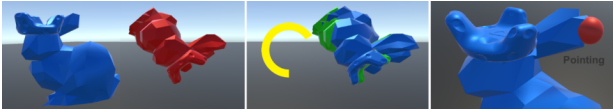

**Figure 6: Object-matching task. The blue bunny (left) is required to match the red's orientation and position. Yellow progress bar and green object color indicate matching. Next, participants point at locations - red sphere on the object.**

The third example, the *crossbow* in Figure 4c, makes use of the linear stretching manipulator with a 40% stretching pattern. The user aims at the target and pulls back on the virtual arrow using the physical block; the virtual crossbow gets rendered with respect to the stretched manipulator. Once, the user lets go, it snaps back and triggers the arrow.

Our last example demonstrates the use of TanGi for controlling a *virtual robotic arm* (Figure 4d). The model for this arm is similar to an existing robotic arm model used in industrial settings [13]. It uses two manipulators, single-axis rotation and linear translation, simultaneously. The user can move the object 6DoF in space; however, they can also rotate the robot wrist independently to adjust and fine-tune the gripper orientation using the single-axis rotation manipulator. To open and close the gripper the physical proxy robotic arm utilizes a linear translation manipulator.

These example applications act as a proof-by-example (as suggested by [33]), and illustrate a wide spectrum of possible use cases for TanGi, from toys and gameplay to industrial applications. To better understand whether composable proxy objects and manipulators provide advantages in terms of usability and naturalness when used for exploration and manipulation, we conducted two user studies.

**STUDY 1: EMBODIED OBJECT EXPLORATION**
Our first study explores how different control types, demonstrating a range of different levels of embodiedness, affect virtual object exploration. Our examples (described above) demonstrate that TanGi does allow building a wide range of proxies for virtual objects, but we wanted to understand the impact of proxies on basic interactions with virtual object (such as reorienting them to get a different view, or interacting with them through natural gestures). To do this, we conducted a controlled lab study where participants re-oriented a virtual object to a pre-specified target orientation and pointed at a target on the virtual object (to represent a simple interaction). Participants compared four different control mechanisms, each with a progressing level of embodiment: (1) free-hand control that approximates natural gesture-based control using a Leap Motion; (2) 6 DoF-controller using a Vive controller; (3) TanGi proxy, which functions as an approximation of the virtual object; and, (4) a high-fidelity 3D print that acts as an exact replica.

### Participants
We recruited 16 participants (seven females; eight males; one preferred not to answer), aged 20-38 (avg: 25.75; sd: 4.5) from the general public and the local university. Participants had a range of different educational and professional backgrounds including engineering, computer science, psychology, chemistry, robotics, music composition, law and modern languages. Two participants had never used VR before, twelve had used it a few times (one to five times a year), one person used it often (6 - 10 times a year), and one other person on a regular basis (more than 10 times a year).

### Procedure
Our study used a within-subjects design, allowing participants to explore and compare the different control types. A Latin-square design was used in order to counterbalance the four condition. The study was conducted in a quiet room to avoid distraction and ensuring the same testing conditions.

After a study introduction and informed consent, participants performed a practice round in VR, giving them an opportunity to familiarize themselves with VR, the study task and the system. When participants felt comfortable, the study began. In the first part of the study, participants were asked to reconstruct the Stanford bunny using the shape primitives available in TanGi. As a reference, a physical 3D printed version of the bunny was provided.

After completing the first part of the study, participants were provided a demographic questionnaire regarding their prior experience and background. Next, they performed a test, to collect data regarding their mental rotation abilities. Finally, they executed the matching task using four different techniques, followed by a final questionnaire as well as a semi-structured interview to better understand their experience. Participants were given a sweet as a token for their participation. The total experiment took approximately 45 minutes, and was approved by the University College London's Ethics Committee.

### Task Design
We chose the 3D printed Stanford bunny for our study, because it has a distinct shape and many details such as ears, tail, nose, etc. Following we describe the two study parts.

*Part 1.* The first part of the study aimed to evaluate the capabilities of our toolkit to approximate and relatively detailed object, and to help us to understand how novices', with no previous experience in this type of proxy creation, approach such tasks. We asked participant to assemble the bunny using our toolkit. There were no constraints given except that the cube with the tracker was required to be the head of the bunny, and therefore was 3D printed with a ¼ inch screw on top. We only offered two different primitive shapes (cubes and half-sphere), each in three different sizes. In our pilot study, we found these shapes were surprisingly sufficient for creating an approximation of the bunny, and put a reasonable cap on the task complexity.

*Part 2*. The task in part 2 models a common operation in a VR world: reorienting an object to locate a particular view and to interact with the object. Our experimental system generates pseudo-random locations on the bunny (red spheres) that indicated where participant needed to find and interact with (through pointing). Subjects were required to alternate between position matching and pointing interactions, and hold a particular position or pointing position for two seconds to complete the task. Figure 6 provides an illustration of the task. Each participant completed ten different orientations and ten pointing locations per condition.

### Apparatus
We implemented the virtual environment in Unity3D [50] (v. 2018.3.11f1) using an HTC VIVE [60] (2PR8100) with SteamVR [61] (v. 1.5.15) and the OpenVR SDK [58] (v. 1.4.18). For the hand tracking we used a Leap Motion sensor [62] (SDK v. 2.3.1) attached to the HTC VIVE. The program was running on a Dell Notebook with an Intel Core i7, 16 GB RAM and NVIDIA GeForce GTX 1060. The ten different locations were hardcoded to ensure that they are always reachable for sitting participants. The required end-locations covered a space of 100x30x25cm (WxLxD) in front of the participant. The pointing locations were randomly selected from a set of five (nose, body, tail, paw and ear of the bunny). To provide support during the task, we displayed a dwell time indicator (during the two-second hold required to complete the task) using a yellow progress bar (see Figure 6). After pilot testing, we chose a rotation threshold of 30 degrees across all three axes, and an overall threshold of 6cm for positioning. Once a participant entered that threshold, the goal bunny turned green and the progress bar started. Participants did not have to select the object in the controller condition. The controller acted as a "stick" for the bunny and its manipulation was immediately displayed on the virtual model.

### Data Collection
We collected data from seven sources: a pre-study questionnaire for demographic information; a mental-rotation test using PsyToolkit [48]; video of the participant as they completed phase one and phase two of the experiment; system logs (including task completion times, accuracy, travelled position/orientation, head movement, head gaze, etc.), field notes and observations, a post-study questionnaire (mainly 7-point Likert scales), and a short semi-structured interview to better understand participants' experiences in the different conditions.

### Analysis
We conducted a statistical analysis (7400 system logged data points), and related this to the results from our thematic analysis where we identified recurring themes in participant behavior as they engaged with the system. In addition, we conducted a modified interaction analysis (Jordan & Henderson [29]) on the videos, where we looked at unusual incidents to provide further insights into how people used the different techniques.

### Results
Here, we show the findings from our two-part experiment. We start with part one where participants were asked to build a rough approximation of the bunny using our toolkit.

*Part 1: Building the proxy object*. All participants successfully assembled a bunny using our toolkit. Two participants reported that it was *"...tricky to match the Velcro tape"* (P11), and suggested that *"...different colors might help"* (P11). However, generally subjects responded positively to *"It was easy to assemble the object."* (md: 6.0; sd: 1.15).

On average, participants took 167s (sd: 44s) to complete the bunny assembly. Participants built 16 different unique bunnies (see auxiliary materials). Participants generally found the shapes they needed, responding to the statement "*All necessary shapes were provided for building the object*" (md: 5.5; sd: 1.93); four participants asked for additional shapes such as triangles. Generally, participants told us that they were satisfied with the result (md: 6.0; sd: 1.59). In responding to why participants built the bunnies the way they did, participants varied on what aspects were more important to replicate. P7 responded *"I just tried to roughly match the size"* whereas P16 stated *"The bunny needs ears!"* showing they wanted to re-create this detail. The toolkit components were largely robust, though the bunny broke towards the end of the second part of the study for two participants and some participants felt uneasy about the stability of bunny appendages while manipulate the proxy—e.g. when the participant included ears. As a consequence, participants were very careful with the proxy, and were gentler when turning it around, as they were afraid the bunny's head would fall off. In spite of this, participants were able to complete all study 2 tasks successfully. As discuss later, alternative construction techniques could address some of these issues, but in general the possibility of breaking a proxy object is a limitation of every block-like construction kit.

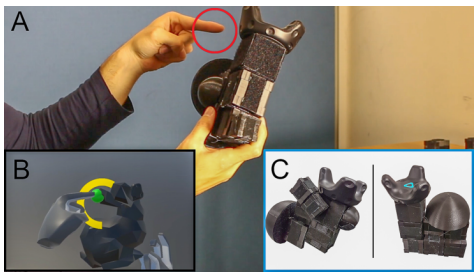

**Figure 7: A/B: Points in the air (red circle), because the object is missing ears. Does not receive tactile feedback. C: Participants built bunny proxies in different shape/size.**

*Part 2: Orienting and Interacting.* All participants used the object they built in Part 1 as the TanGi condition. Our analysis of the mental rotation test did not show any outliers. As illustrated in Figure 8, participants found control types in all conditions easy to learn and easy to use. The 3D printed bunny was the fastest in terms of task completion time. Means for the four conditions were: 3D print (mean: 3.9 s; sd: 1.2 s), Controller (mean: 4.2 s; sd: 1.4 s), TanGi (mean: 5.8 s; sd: 1.7 s), and Free-hand (mean: 10.7 s; sd: 2.3 s).

To further investigate our data, we ran one-way repeated-measures ANOVAs. The collected data sets hold the homogeneity assumption, because they are normally distributed verified through Lilliefors normality tests. Main effects revealed by the ANOVA were tested for significance using post-hoc Bonferroni-Dunn tests.

We found a main effect on task completion times ($F_{4, 45}$ = 130.9, $p < .0001$). Following this, we found a significant difference between Free-hand and the three other conditions as well as between TanGi and the 3D print at $p < 0.05$. This is also supported by participants' ratings to *"I completed the task quickly"* (medians: 3Dprint (6.0), Controller (6.0), TanGi (5.0), and Free-hand (5.0)).

In terms of accuracy we saw similar results. Average error values in degrees across the three rotation axes were: 3D print (mean: 12.7°; sd: 3.1°), Controller (mean: 12.0°; sd: 2.5°), TanGi (mean: 13.7°; sd: 3.3°), and Free-hand (mean: 16.4°; sd: 2.6°). Translation error values along x, y, and z in sum were 3D print (mean: 2.8 cm; sd: 0.7 cm), Controller (mean: 2.7 cm; sd: 0.5 cm), TanGi (mean: 3.2 cm; sd: 0.8 cm), and Free-hand (mean: 3.5 cm; sd: 0.5 cm). We found a main effect for the orientation offsets ($F_{3,45}$ = 20.279, $p < .0001$). Post hoc tests showed a significant difference between Free-hand and the three other conditions at $p < 0.05$. The ANOVA for translation difference indicated a main effect ($F_{3,45}$ = 7.865, $p < .0005$); however, post hoc showed no significant differences after corrections. Participants' ratings align with these findings *"I could orient the object accurately"* (medians: 3Dprint (7.0), Controller (6.5), TanGi (6.0), and Free-hand (5.0)).

*Observations*
**Free-hand**. Without tangible elements it was significantly harder to manipulate the virtual object. We frequently observed that participants were not aware of their grasping point. As with real world objects, the grasping point simultaneously represents the rotation axis. Grasping the bunny at the ear resulted in an unexpected large rotation for participants. In spite of this, two participants favored the virtual condition. *"This is magical…I am not afraid to drop stuff"* (P11) or *"I can just arrange it how I want"* (P16).

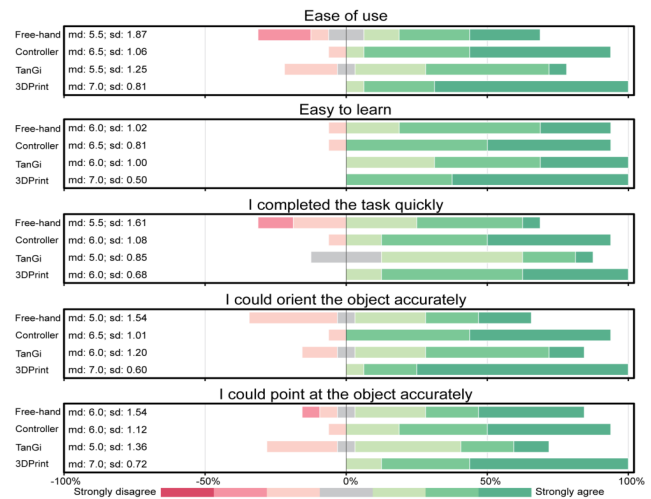

**Figure 8: Post-study questionnaire results on a 7-point Likert-type scale** *(1= Strongly disagree; 7=Strongly agree)*.

**Controller**. The controller provides an easy tangible way to manipulate virtual objects. Subjects reported that it was comfortable to hold and allowed them to easily match the goal orientation. We often observed that rather than changing the grasping position, participants twisted and bended their wrist to rotate the object.

**TanGi Toolkit**. Participants were deliberately slower with the TanGi proxies, as they were worried the components might not stay together. In spite of this, participants performed well using their own proxy. Compared to the Free-Hand and Controller conditions, it allowed them to *"...better understand the size/dimensions of the object"* (P1), *"...because it was closer to what I am holding"* (P10). Participants stated that they used physical parts of the object as landmarks being able to quickly determine the object's orientation: *"I used the tail and the ears so that I roughly know how it is oriented, and it helped me to find the correct pointing location"* (P9). These observations make it clear the proxy functions as an embodied stand-in for the virtual model. This kind of stand-in would be appropriate, opined P8, particularly for *"objects that are challenging to understand in VR, because of the environment, task, rendering, complexity etc. [The proxy] would [allow] my hands to better understand it"* (P8).

One challenge we observed with TanGi proxies was that mismatches between the virtual model and the TanGi proxies caused some confusion. In some cases, we observed that participants overshot the pointing location (i.e. pointed *into* the model rather than on the surface), because they expected to receive tactile feedback about the edge of the virtual model. This would occur, for instance, when parts of the bunny were not replicated in the proxy (e.g. the ears), and tried to touch the tip of the ear. These mismatches slowed participants down, consistent with prior literature [30, 46].

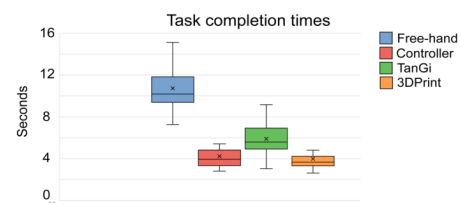

**Figure 9: Task completion times study 1.**

*3Dprint*. The 3Dprint performed best across all measurements, and was also most favored by our participants: *"The 3D print was definitely the best"* (P11) or *"It feels very natural"* (P5). It allowed participants to explore the object, use landmarks to better understand the object and help them especially with the pointing: *"I can just follow the object"* (P13) or *"It allows me to do fine-grain adjustments when I touch it"* (P7). However, four participants told us that they found it challenging to work with the 3D print, because of its size. Furthermore, two stated that they found the weight distribution (center of gravity) confusing. This problem was created by using the HTC VIVE tracker on the head of the bunny, and has also been outlined in previous work [53, 55].

### Study 1 Summary
This study demonstrates that TanGi allows people to build tangible proxy objects that can be used for object exploration in VR. TanGi's proxies helped participant's spatial understanding of virtual objects over the Controller condition, and generally increased their performances compared to free hand interactions. Up to this point, we only investigated how embodied exploration affects user interaction. Therefore, in our second study, we further investigate the use of the manipulators for embodied object manipulation, which bring proxy objects closer to the rich manipulation possibilities of real-world objects.

## STUDY 2: EMBODIED MANIPULATION
While our first study focused on how different control types affected exploration of a virtual object, our second study focused on how embodiment affects manipulations of virtual objects. Specifically, we wanted to understand the impact of TanGi proxies on manipulation tasks. To build this understanding, we conducted a controlled laboratory experiment where participants completed single dimension manipulation tasks. Participants completed trials where each of the three control types (Free-hand, Controller and TanGi) represented a different level of embodied interaction.

We compared our three different conditions in performing three different primitive object manipulations; rotating, stretching and bending. Since, linear translation and linear stretching is essentially the same for the Free-hand and Controller condition, we decided to only include linear stretching in study 2. We did not include a 3D printed condition in study 2, since there is no current analog to 3D printing manipulable objects.

### Participants
We recruited a new set of 12 participants (6 reported as female; 6 reported as male), aged 19-35 (avg: 25.46; sd: 4.8) with a range of professional and educational backgrounds including humanities education, geography, computer science, psychology, environmental science, linguistic, English literature, and civil service. This excludes one participant that was omitted before analysis, due to a problem with experimental system. Each participant was provided a £5 Amazon Voucher as remuneration. Five participants reported that they had never used VR before, five had used it a few times (one to five times a year), and two other subjects use it on a regular basis (more than 10 times a year). Participants from the first study were not permitted to take part in this experiment.

### Task Design
The three tasks were modeled to help us compare three different levels of embodied manipulation: Free-hand, Controller and TanGi. For instance, participants were required to reproduce five different levels of stretch, match five different rotation and bend states within a threshold. Early pilot testing revealed that allowing 6DoF for the Controller and the Free-hand interaction technique was challenging. To ensure the equality of the different conditions we restricted the DoF for Free-hand and Controller. Thus, we essentially implemented a virtual version of the Manipulators by only allowing single axis rotation, linear stretching and unidirectional bending.

Similar to study 1, subjects were required to hold the object for two seconds (indicated through a yellow progress bar). A second object displayed above the one under control showed how much rotation, stretch and bend was required.

### Procedure
After giving participants a general introduction to the study, we explained the task, and showed them the first condition. Next, they performed practice rounds for rotating, bending and stretching, before they did the main experiment. This gave them the opportunity to familiarize themselves with VR, the study task and the condition. We fully counterbalanced the presentation of the three different conditions resulting in six permutations. The experiment took about 30 minutes. The study has been approved by University College London's Ethics Committee.

### Analysis
We followed the same data collection and analysis procedure as in study one.

### Findings & Observations
Here, we focus on the findings from our second experiment. We highlight how people make use of TanGi's manipulators, and we contrast their experiences with the Controller and Free-hand condition to explore embodied manipulations.

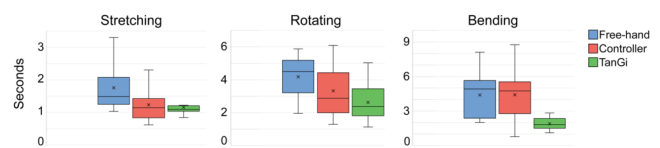

Figure 10: Task completion times study 2.

Overall, TanGi's *Manipulators* outperformed the two other conditions across all measurements. First, we take a look at the task completion times (mean. for one trial) for the three tasks rotation, bending and stretching (also see Figure 10).

***Completion Time***. To further investigate our data, we again ran a one-way RM-ANOVAs after verifying the assumption of normality using a Lilliefors test. Post-hoc tests used Bonferroni-Dunn.

*Rotation:* The times for rotation were Free-hand (mean: 4.18 s, sd: 1.34 s), Controller (mean: 3.32 s, sd: 1.63 s), and TanGi (mean: 2.62 s, sd: 1.17 s). We found a main effect ($F_{2,22}$ = 6.408, $p < .05$). Post-hoc tests revealed that TanGi was significantly faster than Free-hand ($p < .05$).

*Stretching:* For stretching completion times were Free-hand (mean: 1.75 s, sd: 0.72 s), Controller (mean: 1.23 s, sd: 0.49 s), and TanGi (mean: 1.15 s, sd: 0.23 s). The ANOVA showed a main effect ($F_{2,22}$ = 4.429, $p < .05$). We found TanGi to be significantly faster than Free-hand ($p < .05$).

*Bending:* Bending completion times were Free-hand (mean: 4.39 s, sd: 2.06 s), Controller (mean: 4.41 s, sd: 2.18 s), and TanGi (mean: 1.90 s, sd: 0.56 s). A main effect was found ($F_{2,22}$ = 11.969, $p < .05$). Post-hoc tests showed that TanGi was significantly faster than Controller and Free-hand ($p < .05$).

Generally, bending was challenging for participants. Even though we constrained the DoF it still required to manipulate two virtual objects relative to one another. As our early pilot testing showed this confronts participants with challenges.

***Subjective Responses***. Our main analysis aligns with participants' questionnaire responses. For instance, medians for *"Overall impression of the system: I would use the system for virtual 3D object manipulation"* were: TanGi (md: 6.5, sd: 0.52), Controller (md: 6.0, sd: 1.11) and Free-Hand (md: 5.0, sd: 1.54). Participants rated TanGi as *"easy to use"* (md: 7.0, sd: 0.51), compared to Controller (md: 6.0, sd: 1.15) and Free-Hand (md: 4.5, sd: 1.37)); and *"easy to learn"*: TanGi (md: 7.0, sd: 0.28), Controller (md: 7.0, sd: 0.98), Free-Hand (md: 6.0, sd: 1.31)).

Our observations indicated that participants struggled somewhat with the bending task in the Controller and Free-hand condition, which is supported by the completion times. This is also evidenced in the questionnaire responses to *"I could BEND the object accurately"* TanGi (md: 7.0, sd: 0.64), Controller (md: 5.0, sd: 1.50) and Free-Hand (md: 4.0, sd: 1.80). The other tasks (stretching and rotating), which only required the direct manipulation of one virtual object seemed easier. Ratings for *"I could ROTATE the object accurately"* were TanGi (md: 6.5, sd: 0.66), Controller (md: 5.5, sd: 1.37) and Free-Hand (md: 6.0, sd: 1.44); and, *"I could STRETCH the object accurately"*: TanGi (md: 7.0, sd: 0.67), Controller (md: 6.0, sd: 1.19) and Free-Hand (md: 6.0, sd: 0.93). Next, we provide further insights into how people used and experienced the different conditions.

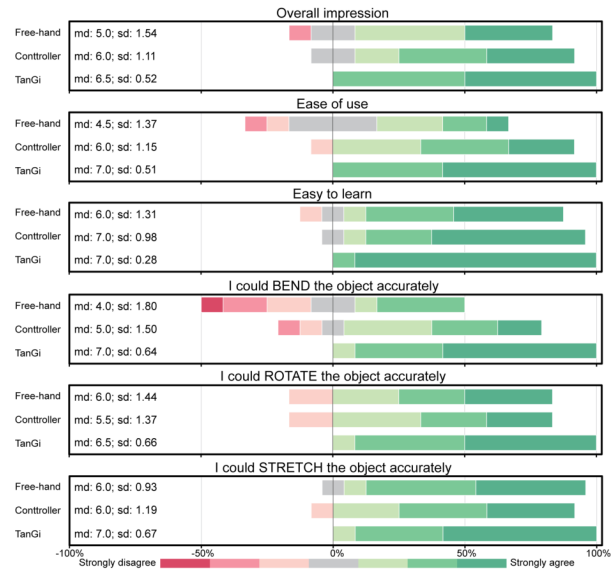

**Figure 11: Questionnaire results on a 7-point Likert-type scale** *(1= Strongly disagree; 7=Strongly agree)*.

***Free-hand.*** Participants had mixed opinions about the free-hand interaction regardless of their prior experience with VR. Performing very specific manipulations required a lot of focused action *"…[I was] very focused on my hand movements"* (P4), because subtle changes in hand orientation was immediately displayed on the object. Interestingly, participants frequently reported that interacting using free-hand was tiring: *"It was very tiring for my arm grasping literally nothing"* (P9).

***Controller.*** The Controller with its uniform shape was slightly preferred over the Free-hand condition, since it provided a tangible way to interact with a virtual model. *"Having an object to hold onto made it easier to keep position of the cubes relative to each other"* (P6). However, some found it cumbersome to use the controller rather than directly interacting with objects as highlighted by participant 11: *"Controller feels like a barrier to the object"*.

***TanGi.*** Overall, TanGi's manipulators performed best offering a *"… direct way to interact with the virtual object"* (P11). Due to the direct mapping between the object interactions (stretch, bend, rotate) people *"…can apply the movement [they] learned"* (P12). Moreover *"[the Manipulators] appeared much easier to stretch, due to the physical feedback (i.e., actually holding two objects in your hand), whereas the other two methods were a little bit more difficult, as they appeared more 'abstract'"* (P7). The Manipulators were treated as if they were the virtual object, but also moved with more care; in contrast, when using either the Controller or the Free-hand, *"I don't really care about the object, I just move it around [until I have completed the task]"* (P2). Finally, the Manipulators allowed users to easily perform *"…subtle adjustments"* (P1), to be very precise and to help participants to *"… better understand the object, its capabilities, and limitations"* (P1).

**Study 2 Summary**

The study demonstrates that TanGi's manipulators enable people to perform complex object manipulations much more easily due to a higher degree of embodiment. Furthermore, it provides interesting insights showing the trade-offs between the different levels of embodiment.

**DISCUSSION & IMPLICATIONS**

Based on our studies, we discuss TanGi proxies and their utility for embodied interaction in VR, identifying opportunities to improve the toolkit.

**Embodied Exploration and Manipulation with TanGi**

The TanGi toolkit gives people the capability to create tangible proxies linked to corresponding VR models. Study 1 showed that people can easily create tangible proxies using the TanGi toolkit. These were good enough for basic exploration tasks such that people's performance with them was on par with a 3D printed virtual object. As we showed in our design explorations and studies, the current prototype of the TanGi toolkit enables a wide range of proxy possibilities.

The tangible proxies enable embodied exploration and embodied manipulation. For the participants in our studies, the proxies were used as if they *were* the virtual object. Exploring different sides of a virtual model and pointing at different parts of it was accomplished by turning the proxy, and pointing at it. Similarly, manipulating different aspects of the virtual model was done by manipulating the proxy. Many participants described developing an understanding of the capabilities and limitations of the virtual model *through* their handling and manipulation of the TanGi proxy. Instead, participants described the Free-hand and Controller conditions as introducing a "layer" between their interactions and the virtual model.

This embodied interaction presents problems when there are mismatches between proxy and virtual model. The tangible proxies are ultimately approximations of the virtual model; as described in Study 1, each participant approximated the bunny in different ways—some built details like ears while others focused on simply approximating size. The problems with the mismatches would manifest in some fairly obvious ways; for instance, participants would overshoot when trying to point/rest their hand on the virtual model's ear if the TanGi proxy did not have ears. Additionally, participants indicated that secondary characteristics of the proxy were also important; for example, the overall *weight* and the *centre of gravity* of the proxy. In Study 1, the TanGi proxy needed to be affixed with a relatively heavy tracker, which threw off how participants expected to be able to handle the proxy (based on how it looked in the VR world). The fact these limitations arose indicate that the TanGi proxies did very much embody the virtual models for participants.

**Improving the Design of the TanGi Toolkit**

While TanGi worked as designed, our experiences provide some clear directions for improvement. TanGi allows people to rapidly build proxies that embody virtual objects by approximating size, shape and manipulations close to what is expected. However, currently TanGi composable blocks are limited in what types of proxies can be created. We believe this can be easily improved upon, for example with additional primitive shapes that few participants asked for. We could easily create a larger range of shapes (e.g. cylinders, pyramids, etc.) in various sizes. This increases the complexity of actually building proxies, but provides more flexibility in the range of models that can be represented.

Furthermore, while we used Velcro to affix blocks to one another, other well-engineered approaches could be leveraged. For example, 3D printed snaps or anchors can be incorporated directly into our 3D prints, providing robust and strong connections that are less likely to break. And while the standard Vive trackers added bulk and weight to the proxies built in Study 1, we could replace them with smaller and lighter emerging trackers (e.g., HiveTracker [23]).

Finally, it might be possible to provide tactile feedback for parts of the proxies that do not have physical manifestation. For example, recent work has shown that worn devices such as temporary tattoos can be used to provide electro tactile feedback [22, 52] or by directly embedding it into the shape primitives [15]. Furthermore, it may be possible to use certain types of haptic retargeting to provide this tactile sensation [6].

**Generalized Controllers with TanGi**

Beyond interacting with VR objects, participants suggested that the TanGi concept could be used for building more generalized, custom input and output controllers as previously shown in [43]. For example, the robotic arm in Figure 4d can be modelled with various manipulators (for steering, rotating and twisting different parts of the arm). In principle, a simple interface to the robot operating system ROS [59] would allow users to control an actual robot arm using TanGi proxies. Other application domains might include AR (e.g. [20, 30] ).

**CONCLUSION**

In this paper we presented TanGi, a toolkit that allows novice users to rapidly build tangible proxy objects in VR. TanGi enables virtual objects to be embodied by approximating their shape and moveable parts, enabling fast and easy virtual object exploration and manipulation. We demonstrated TanGi's flexibility by presenting a variety of potential applications. Through two lab studies we show that different levels of proxy embodiment affect fluidity of virtual object interaction, and that TanGi proxies offer clear advantages over conventional controller. Our work extends the state-of-the-art in virtual reality technology, by demonstrating a new way to build, richer more fully embodied proxy objects.

**ACKNOWLEDGMENTS**

We thank Warren Park and Daniel Gröger for their help with fabrication as well as Tu Dinh Duong, Justas Brazauskas and Ethan Wood for helping us with our studies. Finally, we thank our participants for their time.